\documentclass[sigconf,screen,authorversion]{acmart}

\usepackage{hyperref}
\usepackage[svgnames]{xcolor}
\usepackage{colortbl}
\usepackage{framed}
\usepackage{listings}
\usepackage{xspace}
\usepackage{algorithm}
\usepackage{algorithmic}
\usepackage{makecell}
\usepackage{enumitem}
\usepackage{graphicx}

\setlist[itemize]{leftmargin=1.5em}
\setlist[enumerate]{leftmargin=2em}
\newcommand{\rev}[1]{#1}

\makeatletter
\def\code{%
  \begingroup
  \small\ttfamily
  \@activecode\@code
}
\def\@activecode{%
  \catcode`.\active \catcode`_\active \catcode`'\active \catcode``\active
}
\begingroup\@activecode
\gdef\@code#1{%
  \def.{\char46\allowbreak}%
  \def_{\_\allowbreak}%
  \let'\textquotesingle
  \let`\textasciigrave
  #1%
  \endgroup
}
\endgroup
\makeatother

\newenvironment{rqbox}
  {%
   \begingroup%
   \setlength{\fboxsep}{4pt}%
   \setlength{\fboxrule}{.5pt}%
   \setlength{\parindent}{0pt}%
   \def\FrameCommand##1{\fcolorbox{gray}{gray!15}{##1}}%
   \MakeFramed{\advance\hsize-\width \FrameRestore}%
  }%
  {%
   \endMakeFramed%
   \endgroup%
  }

\lstdefinelanguage{diff}{
    morecomment=[f][\color{Grey}]{@@},
    morecomment=[f][\color{Green}]{+\ },
    morecomment=[f][\color{FireBrick}]{-\ },
}

\newcommand{\ourtool}{PredicateFix\xspace}

\newcommand{\keypredicate}{bridging predicate\xspace}
\newcommand{\keypredicates}{bridging predicates\xspace}

\newcommand{\KeyPredicates}{Bridging Predicates\xspace}
\newcommand{\keyexample}{key example\xspace}

\newcommand{\KeyExample}{Key Example\xspace}
\newcommand{\keyexamples}{key examples\xspace}

\newcommand{\KeyExamples}{Key Examples\xspace}

\newcommand{\figref}[1]{Figure~\ref{#1}}
\newcommand{\tabref}[1]{Table~\ref{#1}}
\newcommand{\secref}[1]{Section ~\ref{#1}}
\newcommand{\algoref}[1]{Algorithm~\ref{#1}}

\copyrightyear{2026}
\acmYear{2026}
\setcopyright{cc}
\setcctype{by-nc-nd}
\acmConference[ICSE '26]{2026 IEEE/ACM 48th International Conference on Software Engineering}{April 12--18, 2026}{Rio de Janeiro, Brazil}
\acmBooktitle{2026 IEEE/ACM 48th International Conference on Software Engineering (ICSE '26), April 12--18, 2026, Rio de Janeiro, Brazil}
\acmPrice{}
\acmDOI{10.1145/3744916.3773159}
\acmISBN{979-8-4007-2025-3/26/04}

\begin{document}


\title[\ourtool: Repairing Static Analysis Alerts with \KeyPredicates]{\ourtool: Repairing Static Analysis Alerts with\\ 
\KeyPredicates}
\renewcommand{\shorttitle}{PredicateFix: Repairing Static Analysis Alerts with \KeyPredicates}


\author{Yuan-An Xiao}
\email{xiaoyuanan@pku.edu.cn}
\orcid{0000-0002-5673-3831}
\affiliation{
    \institution{Key Lab of HCST (PKU), MOE}
    \institution{SCS, Peking University}
    \country{China}
}

\author{Weixuan Wang}
\email{wangweixvan@gmail.com}
\orcid{0009-0005-0309-0990}
\affiliation{
    \institution{Key Lab of HCST (PKU), MOE}
    \institution{SCS, Peking University}
    \country{China}
}

\author{Dong Liu}
\email{liu.dong3@zte.com.cn}
\orcid{0009-0009-3181-5994}
\affiliation{
    \institution{ZTE Coporation}
    \country{China}
}

\author{Junwei Zhou}
\email{zhou.junwei2@zte.com.cn}
\orcid{0009-0002-6546-1740}
\affiliation{
    \institution{ZTE Coporation}
    \country{China}
}

\author{Shengyu Cheng}
\email{cheng.shengyu@zte.com.cn}
\orcid{0009-0005-4541-3305}
\affiliation{
    \institution{ZTE Coporation}
    \country{China}
}

\author{Yingfei Xiong}
\authornote{Corresponding author.}
\email{xiongyf@pku.edu.cn}
\orcid{0000-0001-8991-747X}
\affiliation{
    \institution{Key Lab of HCST (PKU), MOE}
    \institution{SCS, Peking University}
    \country{China}
}

\renewcommand{\shortauthors}{Xiao et al.}

\begin{abstract}
Fixing static analysis alerts in source code with Large Language Models (LLMs) is becoming increasingly popular. However, LLMs often hallucinate and perform poorly for complex and less common alerts. Retrieval-augmented generation (RAG) techniques aim to solve this problem by providing the model with a relevant example, but existing approaches face the challenge of unsatisfactory quality of such examples.

To address this challenge, we utilize the predicates in the analysis rule, which serve as a bridge between the alert and relevant code snippets within a clean code corpus, called \keyexamples. Based on this insight, we propose an algorithm to retrieve \keyexamples for an alert automatically, and build \ourtool as a RAG pipeline to fix alerts from \rev{two static code analyzers: CodeQL and GoInsight}. Evaluation with multiple LLMs shows that \ourtool increases the number of correct repairs by $27.1\% \sim 69.3\%$, significantly outperforming other baseline RAG approaches.
\end{abstract}





\maketitle

\section{Introduction}

Fixing software bugs is a time-consuming job, and many automated program repair (APR) approaches have been proposed in recent decades~\cite{forrest2009genetic,jiang2018shaping,zhu2021syntaxguided,xia2023keep,chen2021sequencer,le2016history}. Among different types of APR approaches, repairing static analysis alerts~\cite{vantonder2018static,bader2019getafix,jin2023inferfix,zhang2023patch,gao2015safe} is of particular interest because static analyzers are widely used in the industry and dealing with their output has become a burden. Companies have integrated APR tools into their developing routine to fix alerts flagged by code analyzers such as Infer and FindBugs~\cite{marginean2019sapfix,kirbas2021introduction}.

Recently, LLMs have shown promising results in various coding tasks~\cite{wang2024software,xia2023automated,jiang2023impact}, and many existing studies have used LLMs to fix static analysis alerts~\cite{jin2023inferfix,berabi2021tfix}. While LLMs are good at fixing many straightforward and common alerts (e.g., unused variables and null pointer exceptions), they suffer from the problem of hallucination and may generate incorrect results for complicated and less common alerts (e.g., uncommon cases of API misuse and complex security vulnerabilities). In particular, we observe that fixing many alerts requires domain-specific or even project-specific knowledge, such as invoking a custom sanitizer or passing a specific parameter. Such knowledge may be absent from the training set of LLMs.

Retrieval-augmented generation (RAG) is a popular technique to improve the performance of LLMs on unfamiliar topics by guiding LLMs with relevant \emph{examples}. Many existing learning-based APR approaches guide the LLM with historical patches~\cite{jin2023inferfix,wang2023rapgen,joshi2023repair} or similar code snippets~\cite{xia2023plastic,zhang2024autocoderover}. However, the quality of such examples limits their effectiveness. It is challenging to find examples highly relevant to the expected patch, particularly for less common alerts where the LLM does not have enough knowledge.

We address this challenge with a novel RAG approach that retrieves example code snippets to guide the LLM to generate a patch. The core problem is how we can know that an example code snippet contains the required knowledge to guide the fix, i.e., a \emph{\keyexample}. We utilize the definition of static analysis rules to address this challenge. Our novel insight is that some predicates in the analysis rule, i.e. \emph{bridging predicates}, can serve as a bridge between the alert and key examples. We will illustrate later that a code snippet is a \keyexample if, by negating a \keypredicate, the alert to fix disappears and a new alert on the example code snippet appears. In other words, the \keyexample contains the code change to fix the alert, as identified by the \keypredicate. We propose an algorithm to identify \keypredicates and \keyexamples.

Based on the above insight, we build a RAG pipeline that collects a corpus of clean code, retrieves \keyexamples in the corpus, and then prompts the LLM with the examples to enhance its repair capability.
We implemented this approach to fix security vulnerabilities as reported by CodeQL~\cite{demoor2008ql}, a popular Datalog-based static code analyzer, \rev{and GoInsight, an internal static code analyzer in ZTE corporation}. We evaluated \ourtool with different LLMs, and the experiment result shows that it can increase the effectiveness of these LLMs by a large number ($27.1\% \sim 69.3\%$), significantly outperforming other baseline RAG techniques.

The contributions of this paper are summarized as follows:
\begin{itemize}
    \item A novel approach to identify key examples for program repair that utilizes the internal state (predicates) of the static analyzer.
    \item An automatic RAG pipeline based on the above approach, which significantly increases the effectiveness of multiple LLMs, as our experimental evaluation shows.
    \item The cross-language dataset consisting of security vulnerabilities in CVE that CodeQL can detect, which is useful in future studies.
\end{itemize}

The remainder of this paper is organized as follows. \secref{sec:motivation} introduces a vulnerability detected by static analysis as a running example to motivate our approach. \secref{sec:approach} illustrates the concept of predicates and bridging predicates, proposes an algorithm to automatically identify \keyexamples based on them, and then describes \ourtool{} -- an end-to-end RAG pipeline that utilizes this algorithm. \secref{sec:impl} describes the implementation of \ourtool on \rev{CodeQL and GoInsight}. \secref{sec:eval} presents the experimental setup and results of our evaluation. \secref{sec:related} compares our approach to related work. \secref{sec:conclusion} concludes this paper.
\section{Motivation and Approach Overview}
\label{sec:motivation}

\subsection{The Running Example}
\label{sec:motivation-example}

\begin{figure*}[tb]
    \centering
    \begin{lstlisting}[language=diff,xleftmargin=5em]
  RMIServerSocketFactory serverFactory = new RMIServerSocketFactoryImpl();
- Map<String, ?> env = Collections.singletonMap(RMIConnectorServer.RMI_SERVER_SOCKET_FACTORY_ATTRIBUTE, serverFactory);
+ Map<String, Object> env = new HashMap<>();
+ env.put(RMIConnectorServer.RMI_SERVER_SOCKET_FACTORY_ATTRIBUTE, serverFactory);
+ env.put("jmx.remote.rmi.server.credential.types", new String[] { String[].class.getName(), String.class.getName() });
  ......
  jmxServer = new RMIConnectorServer(url, env, server, ManagementFactory.getPlatformMBeanServer());
\end{lstlisting}
    \caption{The running example (CVE-2020-13946) and the developer's patch.}
    \label{fig:motivating-example}
\end{figure*}

In \figref{fig:motivating-example}, we use the security vulnerability CVE-2020-13946~\cite{CVE-2020-13946} in the Apache Cassandra database software to motivate our approach. The vulnerability lies in a Java method that misconfigures an \code{RMIConnectorServer} object to allow any type of object for credentials, leaving the chance of executing malicious code through deserialization. The CodeQL code analyzer~\cite{demoor2008ql} can detect this with the rule \code{Insecure\allowbreak{}Rmi\allowbreak{}Jmx\allowbreak{}Environment\allowbreak{}Configuration\allowbreak{}.ql}. It reports an alert to the user about this vulnerability, as shown below:

\begin{rqbox}
\textbf{[Name]} InsecureRmiJmxAuthenticationEnvironment

\textbf{[Severity]} error

\textbf{[Description]} This query detects if a JMX/RMI server is created with a potentially dangerous environment, which could lead to code execution through insecure deserialization.
\end{rqbox}

\figref{fig:motivating-example} also shows the developer's patch for this vulnerability, where the deleted lines are colored red and the added lines are colored green. The patch adds a configuration \code{"jmx.remote.rmi.server.credential.types"} to the variable \code{env} that will allow only a string or an array of strings as the credential, preventing the chance of code execution through insecure deserialization.

The developer's patch indicates that fixing this alert requires knowledge of the specific configuration. Although the configuration is from the official Java SE platform, its usage is sparse in the public domain. On GitHub, searching for this configuration key only returns 87 Java code snippets and 1 Pull Request. As a result, even recent LLMs such as GPT-4o~\cite{gpt4o} and Claude-3.5-Sonnet~\cite{claude} cannot correctly fix this vulnerability. 

The need for such uncommon knowledge is unfortunately common when fixing static analysis alerts. As our evaluation will show, in our benchmark based on 6027 vulnerabilities in the CVE dataset, over 60\% of the alert types appear only once, indicating that they are not frequent in the public domain. Based on the situation in ZTE Corporation, it is also common that an organization or a project defines its own coding rules involving organization- or project-specific knowledge, such as calling a customized sanitizer. LLMs will not learn such knowledge from public training sets.

To enhance the effectiveness of LLMs in this scenario, researchers have proposed various prompt engineering or RAG approaches to provide the model with additional knowledge. However, none of them works for this running example, which we will elaborate on later. In the following subsections, we go through these existing approaches, analyze why they fail, and propose our approach.

\subsection{Existing Prompt Engineering Approaches}
\label{sec:motivation-baseline-basic}

Many existing LLM-based program repair approaches~\cite{xia2023keep,zhang2024critical,wadhwa2024core} prompt the LLM with the natural language description of the alert and the context of the code (typically captured as multiple lines above and below the bug location). However, the vulnerability description in the running example only explains the root cause but does not provide the correct configuration that is critical to the repair. As a result, LLMs still fail to fix this vulnerability with such prompts. For example, GPT-4o would incorrectly remove the original configuration and break the code:

\begin{rqbox}
\textbf{[System]}

You are a helpful programming assistant to fix security vulnerabilities in Java code. You are given a code snippet and a description of the vulnerability. You should briefly describe how to fix the vulnerability, and finally output the modification.

\textbf{[User]}

Vulnerability description: \textit{(alert name)}: \textit{(alert description)}

Code snippet: \textit{(location of the alert with 20 lines of code around)}

\textbf{[Assistant]}

To fix the vulnerability, we need to ensure that the environment map used to create the \code{RMIConnectorServer} does not include any potentially dangerous attributes that could lead to insecure deserialization. Specifically, we should avoid using \code{RMI_SERVER_SOCKET_FACTORY_ATTRIBUTE} directly and instead use a more secure configuration. 

\ldots{}\ldots{}

\begin{lstlisting}[language=diff]
- Map<String, ?> env = Collections.singletonMap(
-   RMIConnectorServer.RMI_SERVER_SOCKET_FACTORY_ATTRIBUTE,
-   serverFactory);
+ Map<String, Object> env = new HashMap<>();
\end{lstlisting}
\end{rqbox}

A possible improvement of the above basic approach is to provide more information, such as the query documentation~\cite{wadhwa2024core}. However, as mentioned, different projects often have different requirements in fixing alerts, e.g., by calling project-specific sanitizers or adhering to project-specific code format, and it is difficult to cover them in a per-rule documentation. 
In this example, GPT-4o follows the instruction in the CodeQL documentation of this rule, which is valid only for Java 10+ programs. It generates \code{env.put(RMIConnectorServer.CREDENTIALS_FILTER_PATTERN, "java.lang.String;!*");}, which is not the correct configuration for this project that targets Java 8+.

\subsection{Existing RAG Approaches}
\label{sec:motivation-baseline-rag}

Retrieval-Augmented Generation (RAG) is a popular technique to enhance LLM by extracting task-related prompts within a corpus of examples. For program repair tasks, a common source of examples is historical patches~\cite{jin2023inferfix,wang2023rapgen,joshi2023repair}. To build such a corpus, approaches typically collect a large set of commits from open-source repositories, pick out potential patches with keywords such as ``fix'' and ``bug'' in commit messages, and run the patches through the static analyzer to determine the alert type that they fix.

However, high-quality historical patches for static analysis alerts are difficult to find~\cite{jain2023staticfixer,zhang2024vuladvisor}. For the running example, there are only 87 code snippets and 1 Pull Request that contain the required configuration key on GitHub. The Pull Request is also a refactor irrelevant to the fix. Therefore, it is nearly impossible to find a patch containing the configuration key without a priori knowledge of its name. As a result, the corpus is often filled with irrelevant changes that dismiss the alert by coincidence (e.g., deleting the function or refactoring to another API), which do not help.

Another possible approach is to search for code snippets similar to the vulnerable code. This approach is widely used in RAG-based~\cite{xia2023plastic,zhang2024autocoderover} and traditional APR approaches~\cite{jiang2018shaping,xin2017leveraging} that do not use LLMs. As identified in existing work~\cite{gao2023what}, BM-25 is a suitable metric to identify similar code snippets for LLMs. However, the precision is often unsatisfactory in our experiment: Although the retrieved code snippets are similar to the vulnerable code, they still lack the key ingredient to fix the alert. For example, below is the retrieved snippet with the top BM-25 similarity to the vulnerable code in the running example:

\begin{lstlisting}[language=java]
  ......
  final MBeanServer jmxServer =
    ManagementFactory.getPlatformMBeanServer();
  try {
    jmxServer.unregisterMBean(new ObjectName(this.getMBeanName()));
  }
  ......
\end{lstlisting}

We see that while some parts (e.g., variable name \code{jmxServer} and method name \code{get\allowbreak{}Platform\allowbreak{}MBean\allowbreak{}Server}) in this retrieved snippet match the code in \figref{fig:motivating-example}, it is semantically irrelevant to the fix, and thus LLMs still fail when prompted with this snippet.

\subsection{Utilizing the Analysis Rule}
\label{sec:motivation-baseline-rule}

The fundamental flaw of the above prompt engineering and RAG baselines is that they rely on high-quality knowledge in the prompts or the retrieved information. Since LLMs generally work well with familiar tasks, and high-quality knowledge is sparse for the other less common tasks (otherwise, such knowledge will be crawled into the training set and the LLM should be familiar with them), these baselines cannot complement the LLM when it fails. A good approach should provide knowledge that is both new to the model and useful to the fix.

An idea to provide such knowledge is to provide the analysis rule. The rule defines whether the analyzer reports an alert or not, so it should include useful knowledge to fix the alert.
In the running example, we can see that the rule includes the exact key name of the configuration (highlighted in pink) inside the \code{putsCredentialtypesKey} predicate, which will dismiss this alert when a correct configuration is present:

\begin{lstlisting}
  module SafeFlowConfig implements DataFlow::ConfigSig {
    predicate isSource(DataFlow::Node source) {
      putsCredentialtypesKey(source.asExpr()) }
    ......
    private predicate putsCredentialtypesKey(Expr qualifier) {
      exists(MapPutCall put |
        put.getKey().(CompileTimeConstantExpr).getStringValue() = [
           <!--\colorbox{magenta!30}{"jmx.remote.rmi.server.credential.types",}->
            "jmx.remote.rmi.server.credentials.filter.pattern"
        ] or ......
      | put.getQualifier() = qualifier and
        put.getMethod().(MapMethod).getReceiverKeyType()
          instanceof TypeString and
        put.getMethod().(MapMethod).getReceiverValueType()
          instanceof TypeObject ) } }
  module SafeFlow = DataFlow::Global<SafeFlowConfig>;
  ......
  from Call c, Expr envArg
  where (isRmiOrJmxServerCreateConstructor(c.getCallee())
      or isRmiOrJmxServerCreateMethod(c.getCallee()))
    and envArg = c.getArgument(1)
    and not SafeFlow::flowToExpr(envArg)
  select c, getRmiResult(envArg), envArg, envArg.toString()
\end{lstlisting}

However, directly prompting the model with the analysis rule does not work either. In the running example, GPT-4o overrides the whole authentication mechanism and leaves it incomplete with a placeholder comment. We recognize three aspects that cause directly prompting with analysis rules suboptimal:

\begin{itemize}
    \item \textbf{Incompleteness:} The analysis rule does not necessarily contain all the ingredients needed to generate a fix. For example, the predicate \code{puts\allowbreak{}Credential\allowbreak{}types\allowbreak{}Key} only contains the name of the configuration key, but not its value. This is reasonable because a program with this configuration key is usually safe. However, without knowing a correct configuration value, we cannot form a patch from scratch. 
    \item \textbf{Identification:} The analysis rule is a long piece of code. In this example, the rule has 89 lines of code and imports more definitions from a shared library. Inlining these definitions leads to even longer code. LLMs may struggle to identify the useful part in such a long input.
    \item \textbf{Understanding:} Static analyzers are only a small and special part of all software on the Internet, and analysis rules are rare in the wild. Therefore, understanding analysis rules is harder for LLMs than understanding regular code snippets.
\end{itemize}

\subsection{Our Approach: Retrieving Examples through the Analysis Rule}
\label{sec:motivation-conditions}

To overcome the above three problems, our approach returns to the RAG paradigm and retrieves a code snippet that demonstrates how to fix the alert for the LLM. In our running example, we would retrieve the following snippet. 

\begin{lstlisting}[language=java]
  ......
    env.put(JMXConnectorServer.AUTHENTICATOR,
      new JMXPluggableAuthenticatorWrapper(env)); }
  env.put("jmx.remote.rmi.server.credential.types", new String[]
    {String[].class.getName(), String.class.getName()});
  return env;
  ......
\end{lstlisting}

Prompting the model with this snippet avoids all the three problems above: (1) it contains complete ingredients, not only the configuration key but also the correct value corresponding to that key; (2) the snippet is short and easy to identify the key part; (3) LLMs are well trained to understand code examples in the same programming language. As a result, models such as GPT-4o or even GPT-4o-mini can generate the correct patch given this snippet.

Now, the problem is how to retrieve such a code snippet. In particular, how do we ensure that the code snippet contains the ingredients needed for fixing, i.e., a \emph{key example}. Since the analysis rule contains some useful knowledge for the fix, we utilize the analysis rule to identify key examples. Concretely, we use a predicate in the analysis rule to relate the alert and the key examples. We call such a predicate a \emph{bridging predicate} as it bridges the gap between the alert and the key examples. Predicates are basic units for determining properties of code fragments, and widely exist in different static analyzers implemented in either logic or imperative programming languages (discussed more in \secref{sec:approach-predicate}), e.g.,  predicate is an explicit language construct in CodeQL. 

Given a reported alert $a$ and a code snippet $s$ from a corpus of clean code, we identify $s$ as a potential key example for $a$ if there exists a predicate $p$ in the analysis rule, such that the following three conditions hold.

\textbf{Condition 1:} Predicate $p$ matches a fragment in $s$, i.e., $p$ evaluates to true on that fragment.

\textbf{Condition 2:} If we negate $p$ in the rule and scan the vulnerable code again, the original alert $a$ should disappear.

\textbf{Condition 3:} If we negate $p$ in the rule and scan the code containing $s$ again, a new alert of the same type should appear.

The first condition ensures that the example contains ingredients relevant to the analysis rule. In the running example, the \code{env.put()} statement setting the configuration key \code{"jmx.remote.rmi.server.credential.types"} is considered as a potential ingredient, because it is matched by the predicate \code{puts\allowbreak{}Credential\allowbreak{}types\allowbreak{}Key}.

The second condition ensures that the example contains ingredients relevant to the current alert \rev{because the predicate matching the example is proven crucial for the alert to appear}. For example, if the rule reports an alert when either predicate A or B is true, and the actual alert instance is based on the predicate A, examples matched by the predicate B are irrelevant to fixing this instance, and will be ruled out by this condition.

The third condition ensures that the example as a whole is relevant to the current alert. Due to the incompleteness of the analysis, predicates that comply with the first two conditions may match code snippets that contain only some part of the relevant ingredients, but are not relevant as a whole. In the running example, using only the first two conditions could identify a snippet that sets the configuration key in another scenario where the configuration value cannot be used in \code{RMIConnectorServer}. 
This condition excludes such examples by ensuring that they also include other parts relevant to the alert (e.g., the \code{new RMIConnectorServer()} statement; otherwise the alert would not appear).

From another perspective, these conditions come from the idea that if $s$ is the intended code change between the vulnerable and patched code, we could theoretically fix the alert by ``adding'' $s$ to the code under repair, or introduce an alert by ``subtracting'' $s$ from the clean code. Due to the differences between the code under repair and the code in the corpus (which may have different variable names and code structures), we could not easily add or subtract a code snippet. However, we can emulate its effect by negating the corresponding predicate $p$. Therefore, these conditions indicate that $s$ is a key example under the abstraction of $p$.

Based on the three conditions, we develop an efficient retrieval algorithm and an automatic RAG pipeline for program repair, as will be discussed in the next section.
\section{Approach}
\label{sec:approach}

\subsection{Predicates in the Analysis Rule}
\label{sec:approach-predicate}
\label{sec:approach-npe-example}

\ourtool assumes a static analyzer is built upon \textit{predicates} -- expressions in an analysis rule that return a Boolean value based on some fragments of the input program. We show that such a concept universally exists in analyzers implemented in either logic or imperative programming languages.

Many static analyzers are implemented in a logic programming language where the concept of predicates naturally exists. For example, Doop~\cite{bravenboer2009strictly} uses Datalog, a popular logic programming language; CodeQL~\cite{demoor2008ql} uses its extended version of Datalog. Below, we give a toy example in Datalog for checking null pointer exceptions (NPEs).

\begin{lstlisting}[language=prolog]
isNull(V, L) :- assignStmt(V, "null", L).
isNull(V, L) :- isNull(V, L0), controlFlowTo(L0, L), !nullGuard(V, L).
nullGuard(V, L) :- assignStmt(V, E, L), constructorCall(E).
nullGuard(V, L) :- assertStmt(V, "!=", "null").
hasAlert(L) :- methodCall(V, _, L), isNull(V, L).
\end{lstlisting}

In a typical Datalog rule for program analysis, there are primitive predicates for matching code snippets in the source code. In this example, \code{assignStmt}, \code{assertStmt}, \code{method\allowbreak{}Call}, \code{constructorCall}, and \code{controlFlowTo} are primitive predicates. 
When a code snippet is matched, a fact is generated to represent this match. For example, predicate \code{assignStmt} matches all assignment statements in the code, and a fact \code{assignStmt(V, E, L)} is generated for each match to represent an assignment statement at the location \code{L} that assigns the value of expression \code{E} to the variable \code{V}. 

Upon a set of primitive facts, an analysis rule generates new facts with inference rules. In the above example, each statement is an inference rule, indicating that the fact on the left-hand side can be generated when all the facts on the right-hand side are present. \code{isNull(V, L)} determines whether the variable \code{V} can be null in the location \code{L}, which holds when either \code{V} is explicitly assigned to \code{null} (line 1) or is previously null in \code{L0} and doesn't reach a null guard at \code{L} (line 2); another predicate \code{nullGuard(V, L)} determines whether location \code{L} can be considered as a null guard of the variable \code{V}, which holds when it assigns \code{V} to a object constructor call \code{E} (line 3) or it asserts the \code{V} not to be \code{null} (line 4). Finally, the \code{hasAlert} predicate prescribes the main outcome of the analysis: If a statement \code{L} is a method call \code{V.anymethod()} where \code{V} can be null, location \code{L} should have an NPE problem. The Datalog engine continues to generate facts using the inference rules until a fixed point is reached. Finally, an alert is reported at location \code{L} for each instance of \code{hasAlert(L)}.
Note that using a location parameter in the predicate is common in analyzers based on logic programming to facilitate error reporting. For ease of implementation, we consider that a predicate matches a code snippet if there exists a fact of the predicate containing a location parameter that points to the code snippet.

Besides logic programming, some static analyzers express analysis rules in the form of imperative programs that loop through each part of the code and check for its properties. For example, Infer~\cite{calcagno2011infer} is implemented in OCaml, and gosec~\cite{gosec} is implemented in Golang. Below is the analysis rule in gosec checking the SSRF vulnerability:

\begin{lstlisting}
  func (r *ssrf) Match(n ast.Node, c *gosec.Context) (*issue.Issue) {
    if <!--\colorbox{magenta!30}{node := r.ContainsPkgCallExpr(n, c, false); node != nil}-> {
      if <!--\colorbox{magenta!30}{r.ResolveVar(node, c)}-> {
        return c.NewIssue(...) } }
    return nil }
  func (r *ssrf) ResolveVar(n *ast.CallExpr, c *gosec.Context) bool {
    if <!--\colorbox{magenta!30}{len(n.Args) > 0}-> {
      arg := n.Args[0]
      if <!--\colorbox{magenta!30}{ident, ok := arg.(*ast.Ident); ok}-> {
        obj := c.Info.ObjectOf(ident)
        if <!--\colorbox{magenta!30}{\_, ok := obj.(*types.Var); ok}-> {
          scope := c.Pkg.Scope()
          if <!--\colorbox{magenta!30}{scope != nil \&\& scope.Lookup(ident.Name) != nil}-> {
            return true }
          if <!--\colorbox{magenta!30}{!gosec.TryResolve(ident, c)}-> {
            return true } } } }
    return false }
\end{lstlisting}

The main procedure of gosec parses the code under analysis and calls the \code{Match} function against each AST node. This function then checks for many properties, each with a Boolean \code{if} condition (highlighted in pink). If all conditions are met, it returns with a \code{NewIssue} call, indicating a reported alert. 

For such analyzers, we consider each Boolean expression used in the \code{if} statement as a predicate, and the code fragment upon which the expression depends as the matched code. The matched code can be identified by a dynamic dependency analysis, or by utilizing the pattern of the analysis rule. For example, the parameter \code{n} in the above example can be directly used as the matched code for the subsequent \code{if} conditions.

\subsection{Matching \KeyExamples}
\label{sec:key-example-conditions}

The three conditions in \secref{sec:motivation-conditions} give us a basic way to determine whether a code snippet is a potential key example or not, but do not allow us to efficiently identify all potential key examples from a large code corpus. Here we propose \algoref{alg:ident-key} as an efficient way of identifying \keyexamples. It takes a target codebase $t$, which has an alert detected by the analysis rule $r$, and also a clean code corpus $C$ with codebases free of alerts detected by $r$.

\begin{algorithm}[tb]
    \caption{Identifying \keyexamples}
    \label{alg:ident-key}





    \begin{algorithmic}[1]
        \REQUIRE Target codebase $t$, analysis rule $r$, clean code corpus $C$.
        \ENSURE \keyexamples $E$.
    
    
        \STATE $P \gets \text{getPredicates}(r)$
        \STATE $P \gets \{p \in P \mathrel{|} \text{checkCond1}(p) \wedge \text{checkCond2}(p, t)\}$ 
        \STATE $E \gets \varnothing$
    
        \FOR{each $c \in C$, $p \in P$} \label{line:ident-key:loop-codebase}
            \FOR{each $s \in \text{getMatches}(p, c)$}
                \IF{$\text{checkCond3}(p, s, c)$}
                    \STATE $E \gets E \cup \{ \text{expandContext}(s) \}$
                \ENDIF
            \ENDFOR
        \ENDFOR
    \end{algorithmic}
\end{algorithm}

The algorithm first initializes a set of possible \keypredicates $P$ as all the predicates appearing in the analysis rule $r$. Then it filters $P$ with the first two conditions: a \keypredicate should match code snippets, and negating it should dismiss the alert. Then, the loop on line~\ref{line:ident-key:loop-codebase} loops through all possible \keypredicates and codebases in the corpus to find possible \keyexamples. We test the third condition for each matched code snippet, and if a snippet passes the test, we expand it to a full key example (by adding several lines of context around the code) and add it to the final result $E$. \rev{In case the static analyzer fails to run due to implementation-level limitations, we skip to the next predicate/example.}

We can see that instead of looping through all possible code snippets and checking each condition individually for each snippet, \algoref{alg:ident-key} checks the three conditions in a particular order (Condition 1, then Condition 2, and finally Condition 3) to optimize efficiency: Checking Condition 1 only requires to parse the argument list of each predicate, which does not depend on the corpus and takes nearly no time; checking Condition 2 requires to re-run the analysis on the code under repair, but does not depend on the corpus; checking Condition 3 requires to run the analysis \rev{for each project in the corpus, which is computationally heavy if there are a large number of projects, so we put it at the last step.}

Since there are many different ways to write a rule, the three conditions are heuristic and cannot guarantee that all identified examples are key to the fix. For example, the rule may capture two distinct patterns for an alert, where the buggy code falls into one pattern, the example falls into the other pattern, and the bridging predicate is used in both patterns. The prioritization step that we will describe in \secref{sec:example-prio} is designed to deal with such cases. 

\subsection{The RAG Pipeline}

Finally, we discuss how to build a RAG-based program repair tool, \ourtool, based on \algoref{alg:ident-key} that finds \keyexamples.

\begin{figure}[tb]
    \centering
    \includegraphics[width=.9\linewidth]{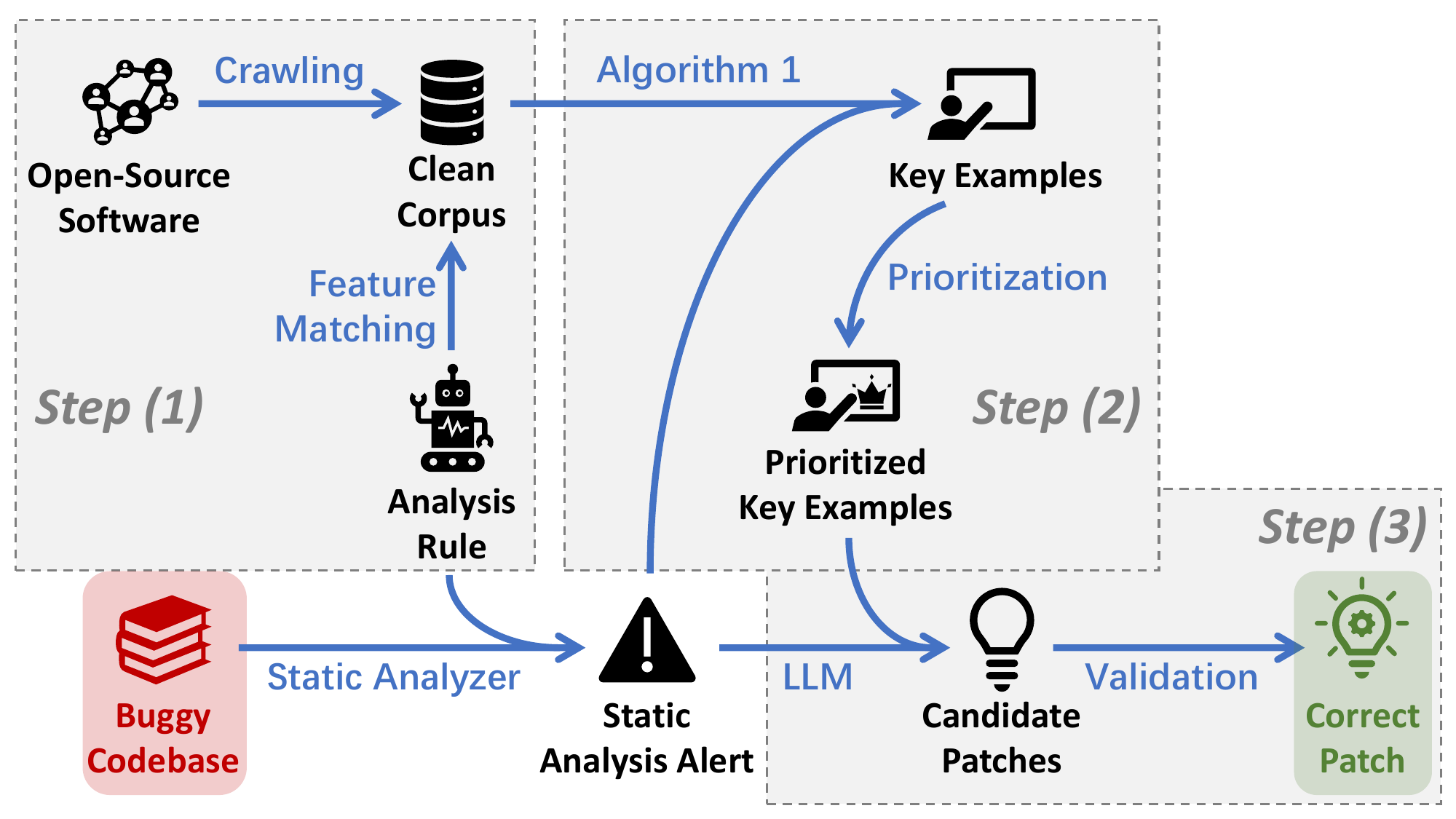}
    \caption{\rev{The RAG pipeline of \ourtool.}}
    \label{fig:approach-overview}
\end{figure}

The overview of \ourtool is shown in \figref{fig:approach-overview}. \rev{Given that the analysis rule reports an alert on the target codebase, \ourtool takes three major steps to fill the gap between the alert and the fix: (1) It collects a corpus of clean code based on various sources; (2) it runs \algoref{alg:ident-key} to get \keyexamples from the corpus, and then prioritizes \keyexamples to form the prompt with the most relevant \keyexamples; (3) it calls the LLM to obtain candidate fixes and validates each fix against the analysis rule.
In the following, we discuss each important part of this pipeline categorized by its step.
}

\subsubsection{Corpus Collection}
\label{sec:approach-corpus}

Any RAG approach requires a corpus from which the examples are retrieved. The completeness of the corpus is essential in the whole approach, because there is no way to retrieve a relevant example in case it is absent in the corpus. We build a corpus containing mostly clean code from three sources.

\begin{itemize}
    \item The first source is \textbf{open-source software repositories} in the same programming language as the target codebase. We can assume that their code quality is good enough, but they may lack \keyexamples using some less common APIs or features.
    \item To augment the corpus when the analysis rule is less common, we additionally search for \textbf{repositories matching string literals} in the analysis rule or code examples in the documentation of the analysis. For example, when fixing the running example as described in \secref{sec:motivation}, we recognize the literal string \code{"jmx.remote.rmi.server.credential.types"} in the analysis rule and search for relevant code repositories.
    \item We can also add \textbf{other parts of the target codebase} or other \textbf{user-specified additional codebases} to the corpus. This is because the user may have some distinct code styles and patterns, and we may get a relevant example in these codebases.
\end{itemize}

\rev{Since the corpus only depends on the analysis rule instead of the alert, users could practically collect and cache the corpus for each analysis rule before repair. This can improve the efficiency of \ourtool by offloading some analysis time.}

\subsubsection{\KeyExample Prioritization}
\label{sec:example-prio}

According to \citet{gao2023what}, more examples may not always lead to better LLM performance, and their experiment has shown that four examples are enough for bug fixing. We take a gradual approach to minimize the cost and response time of \ourtool: It iteratively calls the LLM, starting with no example (same as the Basic baseline in \secref{sec:motivation-baseline-basic}), and picking the next example if the LLM does not repair successfully. It stops when a patch fixing the alert has been found or all of the first four examples have failed for each source of the corpus.

We apply several prioritization rules to the \keyexamples from the list returned by \algoref{alg:ident-key}, as shown below. 

\begin{itemize}
    \item \textbf{Similarity heuristic:} We calculate the BM-25 similarity between each \keyexample and the code context of the alert. The examples with the highest similarity are ranked at the top. Since BM-25 requires the code to be tokenized, we use the BPE tokenizer from CodeLlama~\cite{roziere2024code}, which is trained with source codes and can handle out-of-vocabulary identifier names.
    \item \textbf{Count heuristic:} We count the number of occurrences of each \keypredicate in the \keyexamples list. If a \keypredicate has too many occurrences, we drop \keyexamples matched by this predicate or rank them at the bottom. The rationale for this heuristic rule is that such predicates are typically not specific enough to be useful (e.g., matching all constructor calls or all assignment statements).
    \item \textbf{Code hierarchy heuristic:} If the file containing a \keyexample or its \keypredicate is located in a library path, we drop the example or rank it at the bottom. This is because such examples are not project-specific (for \keyexamples) or rule-specific (for \keypredicates), so they are less likely to be fix ingredients.
\end{itemize}

\rev{These rules improve the quality of \keyexamples and the efficiency when there are false positives (irrelevant predicates/examples) matched by the three conditions in \secref{sec:motivation-conditions}. For example, when fixing \code{XssThroughDom.ql} alerts in CodeQL, a predicate matches all \code{length} field access for string variables. Therefore, all \code{xxx.length} expressions would be incorrectly considered as key examples. These false positives will be ruled out by the count heuristic.}

\subsubsection{Interfacing with LLMs}
\label{sec:output-format}

Large language models receive input and give output in the form of natural language conversation, so we need to organize relevant information to fix the alert in a natural language format (``prompt''), and instruct the model to respond with the patch in a format that \ourtool can automatically parse and validate. The quality of the input/output formats may influence their effectiveness.

Our prompt format generally follows the OpenAI guide~\cite{openai_prompt_guide} and contains the task description, the output formatting instruction, the natural language description of the alert, the context of the code to repair, and retrieved \keyexamples. The task description and the output formatting instruction are in the system prompt, and other task-specific contexts are in the user prompt, as shown below.

\begin{rqbox}
\textbf{[System]}

You are a helpful programming assistant to fix security vulnerabilities in \textit{(language)} code. You are given a code snippet and a description of the vulnerability. You should briefly describe how to fix the vulnerability, and finally output the modification in JSON.

In the JSON, for each modification, use \textasciigrave old\_line\textasciigrave{} to mark the exact line to modify, and \textasciigrave new\_line\textasciigrave{} as the modified line. Example format:

\begin{lstlisting}[language=java]
```json
[{ "old_line": "int x = 1;",
   "new_line": "int x = 1; x++;" }]
```
\end{lstlisting}
\end{rqbox}

\begin{rqbox}
\textbf{[User]}

Vulnerability description: \textit{(alert name)}: \textit{(alert description)}

Code snippet: \textit{(location of the alert with 20 lines of code around)}

Below code snippet is a safe example. You can use it if helpful.

\textit{(the \keyexample with 6 lines of code around)}
\end{rqbox}

Note that the output format instructed in the prompt is different from the standard GNU diff format. This is because models often generate invalid diffs~\cite{liu2024marscode}, such as generating wrong line numbers (\code{@@ -1,1 +1,1 @@}) and forgetting to put a space at the beginning of each unmodified line. As a result, these diffs cannot be parsed by machines. In contrast, recent models are fine-tuned to output valid JSON, so this output format leads to better effectiveness.

After receiving the response from the LLM, we extract the JSON code block containing the patch and then apply each of the line modifications. The patched codebase is then re-analyzed by the same analysis rule, and if the alert disappears, the fix is considered successful and provided to the user as the final patch. \rev{Note that a \textit{successful} patch that passes this check may not be the \textit{correct} one wanted by the user. For example, the LLM often generates placeholder or function-breaking codes. Human inspection is necessary to determine the correctness of patches.}
\section{Implementation}
\label{sec:impl}

\subsection{On Logic Programming Analyzers}

We have implemented \ourtool to fix the security vulnerabilities reported by the CodeQL~\cite{demoor2008ql} static analyzer, a popular static analyzer integrated in GitHub~\cite{github_code_scanning} and used by many open-source projects such as Chromium~\cite{chromium_codeql}. We chose this static analyzer as our target because it shows the ability to detect non-trivial security vulnerabilities across different programming languages, while many other static analyzers only contain a few types of analyses or only report trivial issues such as unused variables. 
The rules of CodeQL are also open source, which facilitates our implementation.

\textbf{Corpus selection:} \secref{sec:approach-corpus} discusses different sources of the clean code corpus. For the first source (popular software), we search for the top 1000 repositories in the programming language of the target codebase, and then clone them from GitHub. We also include other repositories in the benchmark in the corpus. For the second source (matching string features), we use the code search feature in GitHub and clone the top 3 repositories. \rev{This hyperparameter is determined by our experience with a few rules: A good key example is generally within the first few search results if the keyword is specific enough (e.g., \code{jmx.remote.rmi.server.credential.types}); otherwise, the keyword is likely too general (e.g., \code{toString}) so keeping more results will not improve data quality.} We do not assign other additional codebases as the third source. To avoid possible data leakage in the corpus, we drop the code from the corpus if it precisely matches the developer's patched code.

\textbf{Example prioritization parameters:} \secref{sec:example-prio} introduces three heuristic rules to filter and prioritize \keyexamples. For the count heuristic, we remove a \keyexample if the number of occurrences of the corresponding \keypredicate is greater than 20. For the code hierarchy heuristic, we remove \keypredicates located in basic language definition rules such as \code{Expr.qll}. For the similarity heuristic, we pick the top four \keyexamples with the highest BM-25 similarity, following the existing paper~\cite{gao2023what}.

\subsection{On Imperative Analyzers}
\label{sec:impl-goinsight}

To evaluate whether our approach works for static analyzers implemented in different programming paradigms (i.e., logic or imperative), we applied \ourtool on GoInsight, an internal Golang code scanner in ZTE Corporation. GoInsight is an imperative static analyzer with 91 analysis rules written in Golang. The implementation of this analyzer is similar to gosec~\cite{gosec}, as we have discussed in \secref{sec:approach-predicate}. The analysis rules are customized to be more accurate and cover more types of issues than the built-in rules in gosec.

\textbf{Predicate extraction:} \algoref{alg:ident-key} requires a mechanism to obtain and negate predicates behind an analysis rule. For this purpose, we modified the source code of GoInsight to add a trace statement to each Boolean calculation, such as an \code{if} condition. We can then negate a specific predicate corresponding to a trace ID in the source code. This allows \ourtool to control the execution process of the analyzer, including obtaining and negating these predicates.
\section{Evaluation}
\label{sec:eval}

In this section, we empirically evaluate \ourtool with the following research questions.

\begin{enumerate}[left=1em]
    \item[RQ1.] \textbf{Effectiveness:} Can \ourtool increase the effectiveness of LLMs in fixing static analysis alerts?
    \item[RQ2.] \textbf{Comparison with other techniques:} Is \ourtool better than other possible retrieval-augmented generation (RAG) and prompt engineering techniques?
    \item[RQ3.] \textbf{Generalization:} How does the effectiveness of \ourtool generalize to another static analyzer?
\end{enumerate}

\subsection{Experimental Setup}

\subsubsection{The Benchmark}
\label{sec:eval-benchmark}

\textbf{RQ1-2} evaluate the effectiveness of fixing security vulnerabilities reported by CodeQL~\cite{codeql_rules}.

We first constructed a dataset of security vulnerabilities. There are multiple existing datasets, including BigVul~\cite{fan2020code}, CrossVul~\cite{nikitopoulos2021crossvul}, Vul4J~\cite{bui2022vul4j}, and Project KB~\cite{ponta2019msr}, all of which are based on the Common Vulnerabilities and Exposures (CVE) dataset. We merged these datasets into a cross-language dataset containing 6027 CVE entries with the Git repository of the source code and the exact commit ID of the developer's patch.

Then, we ran all of the built-in analysis rules in CodeQL against all entries in Java, JavaScript, Python, or Ruby. We chose only these four languages because analyzing programs in other languages (e.g., C++ and Go) requires an environment to compile and build the codebase, which cannot be set up automatically. For each entry, we compare the analysis results before and after the the developer's patch commit. If one or more alerts are eliminated, the commit is marked as fixing that CodeQL alert and included in the benchmark. We further manually inspected these commits and discarded 13 commits that made an irrelevant change and unintentionally caused the alert to disappear (e.g., by refactoring to another API not covered by CodeQL). The final benchmark for our experiment contains 117 verified vulnerabilities that CodeQL can recognize with the corresponding patches from developers. It includes 67 vulnerabilities in JavaScript, 33 in Java, 10 in Python, and 7 in Ruby.

\begin{figure}[tb]
    \centering
    \includegraphics[width=.95\linewidth]{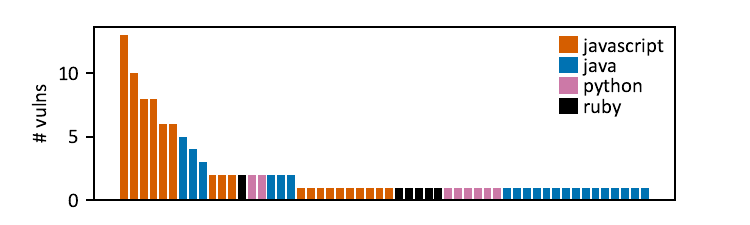}
    \caption{Distribution of 117 CVE vulnerabilities detected by 53 CodeQL rules.}
    \label{fig:eval-setup-distribution-codeql}
\end{figure}

These 117 vulnerabilities are detected by 53 CodeQL rules following a long-tail distribution, as shown in \figref{fig:eval-setup-distribution-codeql}. We can clearly see the distinction between different types of vulnerabilities: While the most common rule (\code{XssThroughDom.ql}) flagged 13 JavaScript vulnerabilities, many other vulnerabilities are less common, including 34 rules that recognize only one vulnerability in the whole CVE benchmark, taking up 64\% of rules or 30\% of alerts.

\begin{table}[tb]
    \centering
    \caption{\rev{Distribution of 416 alerts detected by 91 GoInsight rules. 17 discarded alerts are not suitable for automatic repair.}}
    \begin{tabular}{lrr}
        \toprule
        Issue Type           & \#Rules & \#Alerts (+Discarded) \\
        \midrule
        \code{BUG.*}         & 36      & 142 (+11) \\
        \code{PERFORMANCE.*} & 39      & 179 (+6) \\
        \code{SECURE.*}      & 16      & 78 (+0)   \\
        \midrule
        $\Sigma$ Total       & 91      & 399 (+17) \\
        \bottomrule
    \end{tabular}
    \label{tab:eval-setup-distribution-goinsight}
\end{table}

\textbf{RQ3} evaluates the generalization of our approach \rev{by applying \ourtool on GoInsight, as discussed in \secref{sec:impl-goinsight}. GoInsight is an imperative code analyzer that covers different types of issues, as shown in \tabref{tab:eval-setup-distribution-goinsight}. Among the 91 rules in total, 36 rules check for logical bugs in the program, such as missing switch cases and dereferencing null pointers; 39 rules check for performance issues, such as inefficient string comparison and unused assignment statements; 16 rules check for security vulnerabilities, such as using weak hash algorithms and possible integer overflows during conversion.

To build the benchmark of RQ3, we retrieved the list of Golang repositories with the most stars on GitHub. For each repository, we then ran GoInsight against it and collected alerts with the corresponding alert-triggering code snippets for each rule. False alarms are skipped in this process through manual inspection. We kept at most five alerts per rule to avoid flooding the benchmark with too many simple alerts. The process continued until each rule had at least one alert.
The collected test suite contains 416 verified alerts in total: 5 alerts for each of 78 rules, 1-4 alerts for each of the remaining 13 rules. We further discarded 17 alerts that are not suitable for an automatic repair (e.g., a rule warns about empty loop bodies, but such bodies are intentionally left empty in some code, so the intended repair is not well-defined). Finally, we got 399 remaining alerts as the benchmark for RQ3, covering 346 projects.
}

\subsubsection{Language Models}

For \textbf{RQ1-2}, our evaluation included a variety of LLMs from different vendors:

\begin{itemize}
    \item GPT-4o (2024-08-06) and GPT-4o-mini (2024-07-18) from OpenAI~\cite{gpt4o};
    \item DeepSeek-V3 (0324) and DeepSeek-R1 (0528) from DeepSeek~\cite{deepseekai2025deepseekr1};
    \item Claude-3.5-Sonnet (20241022) from Anthropic~\cite{claude};
    \item Qwen2 from Alibaba Group~\cite{yang2024qwen2}.
\end{itemize}

\rev{The LLMs cover both commercial and open-weight models from different vendors, and also cover both best-performing and cost-efficient models. Therefore, we believe that the selection can represent the recent development of LLM at the time the experiment is conducted (in January 2025).}

Following existing papers~\cite{gao2023what,li2024llmassisted} and OpenAI's guide~\cite{openai_prompt_guide}, the temperature parameter is set to zero for all models, which is optimal for coding tasks and helps reproduce the experiment.

\rev{For \textbf{RQ3}, we used the internal LLM deployed in ZTE Corporation, which powers various code intelligence functions for the employees.}

\subsubsection{Baselines Approaches}

For \textbf{RQ1} and \textbf{RQ3}, we compare \ourtool against the Basic baseline approach as described in \secref{sec:motivation-baseline-basic}. The Basic approach prompts the model with the alert name, its natural language description, and the code context, without additional RAG techniques. Therefore, it can be estimated as the intrinsic effectiveness of that LLM. \ourtool will be empirically helpful if its effectiveness is greater than the Basic baseline. 

For \textbf{RQ2}, we consider three possible RAG techniques, as listed below. The prompts used in all these baselines include the alert details (as in the ``Basic'' baseline) and additional retrieved information.

\begin{itemize}
    \item ``History'' (described in \secref{sec:motivation-baseline-rag}), which retrieves the historical patch identified in the wild;
    \item ``Similar'' (described in \secref{sec:motivation-baseline-rag}), which retrieves the most relevant code snippets with BM-25 similarity;
    \item ``Rule Src'' (described in \secref{sec:motivation-baseline-rule}), which retrieves the source code of the analysis rule, truncated to 1000 tokens;
\end{itemize}

We also consider two better prompt engineering techniques:

\begin{itemize}
    \item ``Rule Expl'', which enhances ``Rule Src'' by asking the model to first explain the source code in natural language, and then fix the bug given the explanation;
    \item ``Docs'' (described in \secref{sec:motivation-baseline-basic}), which enhances ``Basic'' by providing the full CodeQL documentation (i.e., the \code{.qhelp} file) of the alert.
\end{itemize}

We compare the increase in effectiveness of \ourtool with each of the five techniques above.

\subsubsection{The Metric}
\label{sec:eval-setup-succ-def}

We consider a repair to be \textit{successful} if the model output contains a valid patch, and the alert from the static analyzer disappears after applying the patch.
However, a successful repair might not be \textit{correct}: a patch deleting the entire vulnerable functionality certainly causes the alert to disappear (thus successful), but it is highly unlikely to be the repair expected by the developer (thus incorrect). Human inspection is necessary to distinguish correct repairs from plausible repairs.

\rev{Therefore, for \textbf{RQ1} and \textbf{RQ3}, we manually inspected the patches to flag successful patches that are incomplete code, break existing functionalities or introduce new issues to be incorreect. After this process, we report the number and percentage of correct repairs, which is a more suitable metric than successful repairs.}

Only for \textbf{RQ2}, we cannot perform a human inspection because the total number of patches is large (up to 117 alerts $\times$ 6 models $\times$ 5 new baselines). Therefore, we only report the number of successful repairs in RQ2. We argue that: (1) Manual inspection on RQ1 will show that 85\% successful repairs are correct, so the precision of the successful metric will be adequate for a qualitative finding given the results in RQ2; (2) it is common in existing publications~\cite{ramos2024batfix,lutellier2020coconut} to rely on an automatic metric or to inspect only on a small set of patches when a full human inspection is impossible;

The evaluation does not discuss the time used to fix the vulnerability because the response time of LLMs is highly unstable depending on the service load. Since the static analysis result and intermediate data can be cached for the corpus, most of the work in \ourtool can be done in advance once per rule. 
Therefore, the impact of \ourtool on time usage can be minimal (within one minute per bug in most cases if properly optimized).

The evaluation also does not discuss the costs of calling LLMs because a key example is generally a few lines of code consisting of at most several hundreds of tokens. Therefore, the token cost is not a severe limitation of \ourtool or any other baseline. In fact, the whole experiment costs us less than \$50 to fix 117 alerts with the 6 LLMs and 7 approaches (\$0.01 per alert on average).

\subsection{RQ1: Effectiveness}

To evaluate the effectiveness of \ourtool, we count the number of successful and correct repairs with \ourtool against the Basic baseline for each LLM. The results are shown in \tabref{tab:eval-effectiveness}.

\renewcommand\theadfont{}
\begin{table}[tb]
    \centering
    \caption{Number of correct (successful) repairs against 117 CVE vulnerabilities with various LLMs.}
    \label{tab:eval-effectiveness}
    \begin{tabular}{lccc}
    \toprule
        Model & \thead{Basic \\ Baseline} & \thead{With \\ \ourtool} & Improvement \\
    \midrule
        GPT-4o & 59 (70) & 75 (92) & 27.1\% (31.4\%) \\
        GPT-4o-mini & 49 (60) & 66 (84) & 34.7\% (40.0\%) \\
        DeepSeek-V3 & 47 (52) & 76 (88) & 61.7\% (69.2\%) \\
        DeepSeek-R1 & 51 (58) & 74 (84) & 45.1\% (44.8\%) \\
        Claude-3.5-Sonnet & 56 (62) & 81 (91) & 44.6\% (46.8\%) \\
        Qwen2 & 46 (55) & 73 (85) & 58.7\% (54.5\%) \\
    \bottomrule
    \end{tabular}
\end{table}

We can see that for all six LLMs, \ourtool improves the number of correct repairs by 27.1\% to 61.7\%, or successful repairs by 31.4\% to 69.2\%. This result indicates that \ourtool steadily increases the effectiveness of LLMs in the program repair task.

\rev{From the numbers in \tabref{tab:eval-effectiveness}, \ourtool increases the number of successful patches by 167, while increasing the number of correct patches by only 137. The difference between these two numbers (30 patches) indicates the cases where we generate a plausible (successful but incorrect) patch. Overall, \ourtool slightly increases the plausible rate among successful patches from 13.7\% to 15.0\%. However, Since the increase in correct patches is significant, we believe that \ourtool is overall beneficial to users.}

With the help of \ourtool, the best model (Claude-3.5-Sonnet) can correctly repair 81/117 = \textbf{69.2\%} bugs in this benchmark; even the worst model (Qwen2) can now repair 73/117 = \textbf{62.4\%} bugs, significantly better than any LLM in the Basic baseline. Therefore, \ourtool opens an opportunity for program repair users to switch to a faster and cheaper model while the effectiveness is even increased. For example, GPT-4o-mini with \ourtool correctly fixes 7 more bugs than GPT-4o with basic prompt engineering.

\rev{During the manual inspection of patches, we found \ourtool especially helpful in two scenarios: (1) Some vulnerabilities have multiple directions to fix, with varying difficulties. LLMs may stick to a difficult direction and fail to generate a fully correct patch in the end. With multiple key examples, LLMs can be better aware of other possible directions. (2) When the fix involves a particular configuration, LLMs may be unsure about the exact configuration to add, leaving a placeholder for the user to fill in (e.g., \code{// implement access control here}). With relevant key examples, LLMs can better generate the complete configuration.}

Another interesting observation is that the performance of Deep\allowbreak{}Seek-R1 is on par with DeepSeek-V3, contradicting the common belief that a reasoning model should be better than the base model. A possible explanation is that the bottleneck for the current task is the awareness of specific knowledge instead of the reasoning ability. If the model is not aware of an important fix ingredient for a less common alert type, it cannot correctly fix the alert anyway, with or without reasoning.

\begin{rqbox}
\textbf{Finding 1:}

\ourtool can significantly improve the effectiveness of LLMs to fix static analyzer
alerts, increasing the number of correct repairs by a large number ($27.1\% \sim 61.7\%$). The increase is observed in all six LLMs studied in this RQ.
\end{rqbox}

\subsection{RQ2: Comparison with Other Techniques}

To compare \ourtool with other possible RAG and prompt engineering techniques, we implemented the other five baselines described above and compared the number of successful repairs. We further break down the number into each programming language to measure the steadiness of the improvement. 

\begin{figure}[tb]
    \centering
    \includegraphics[width=.88\linewidth]{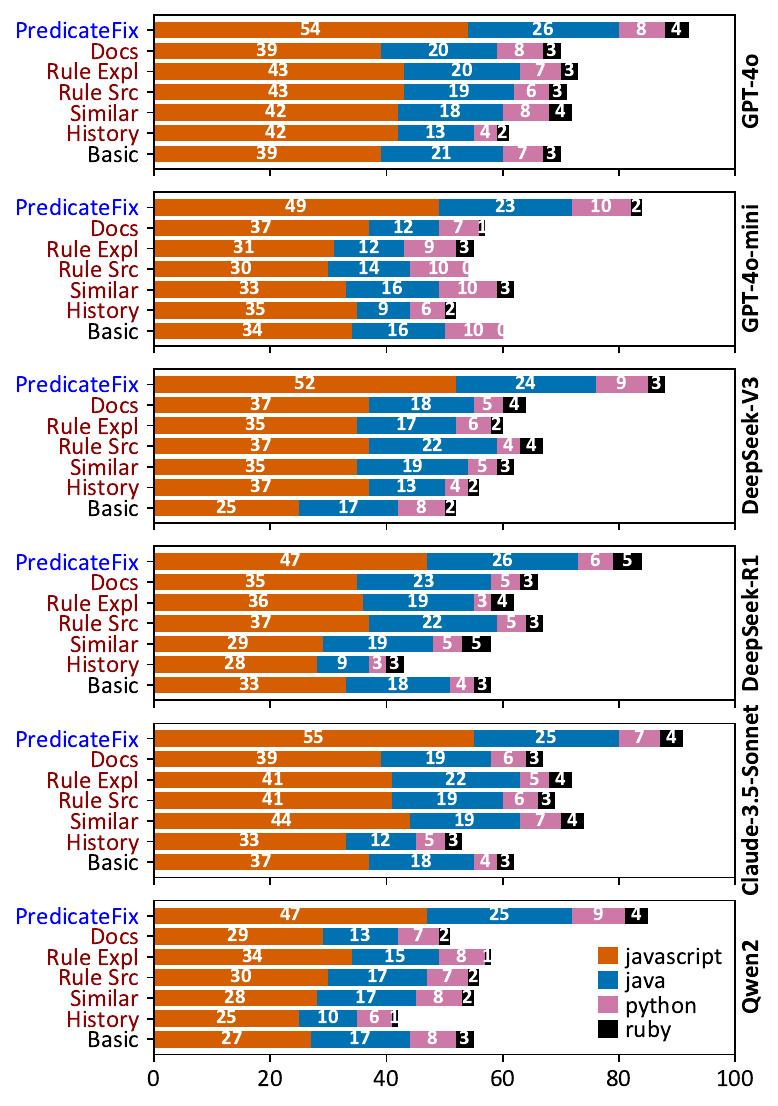}
    \caption{Number of successful repairs against 117 CVE vulnerabilities between \ourtool and baselines.}
    \label{fig:eval-baselines}
\end{figure}

From \figref{fig:eval-baselines}, we can see that the five new baselines (labeled in dark red) cannot steadily increase the effectiveness over Basic. Their increases are small and vary across different LLMs, and can even be negative for some models (e.g., GPT-4o-mini and Qwen2). We speculate that these models have a weaker ability to judge the relevance of the provided RAG information, and they may be misled when the information is irrelevant to the correct fix. In contrast, \ourtool provides the models with high-quality key examples and thus significantly outperforms the baselines. \rev{As a result, \ourtool generates unique fixes that all baselines fail, including the running example discussed in \secref{sec:motivation-example}. For each LLM in \figref{fig:eval-baselines}, the number of unique fixes is 4, 5, 4, 3, 2, and 7.}

Furthermore, the effectiveness increase of \ourtool is steady in different programming languages and alerts, indicating that it helps the LLM to be more effective even for common alerts. This can be seen as a by-product of the proposed RAG pipeline: a variety of examples can instruct the model to think in multiple aspects, hence further increasing its effectiveness on tasks it already has relevant knowledge about.

\begin{rqbox}
\textbf{Finding 2:}

\ourtool can steadily increase the effectiveness of repair \rev{over the Basic baseline, with multiple unique fixes}. In comparison, the other five RAG and prompt engineering baselines have a weaker or even negative increase for some LLMs.
\end{rqbox}

\subsection{RQ3: Generalization}

\begin{figure}[tb]
    \centering
    \includegraphics[width=1\linewidth]{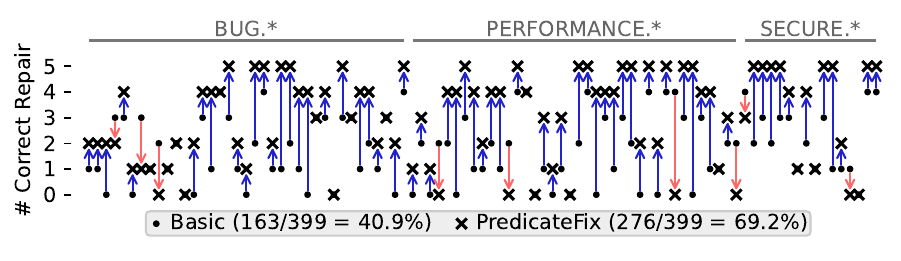}
    \caption{\rev{Correct repairs for each GoInsight alert type. Blue arrows indicate that \ourtool increases the number of correct repairs. Red arrows indicate a decrease.}}
    \label{fig:eval-rq3}
\end{figure}

\rev{
To evaluate the generalization of \ourtool beyond fixing vulnerabilities detected by CodeQL, we applied it to GoInsight to form a repair system for general Golang bugs. We also combined GoInsight with basic prompt engineering as the Basic baseline. We then counted the number of correct repairs of these two systems against the benchmark described in \secref{sec:eval-benchmark}.

On the Basic baseline, the system has generated \textbf{163} correct repairs in total. Breaking down into categories, the number consists of 59 correct repairs for \code{BUG.*} alerts, 69 for \code{PERFORMANCE.*} alerts, and 35 for \code{SECURE.*} alerts. During manual inspection, we found that LLM often hallucinates while repairing, possibly because Golang is a new programming language. For example, to fix the \code{BUG.NO.SWITCH.CASE} alert, we need to add the missing cases in a \code{switch} clause, and the missing cases to add are already instructed by the static analyzer. The repair should be an easy task, but in one case, the LLM ignores the instruction and determines that a \code{break} statement should be added at the end of each switch case. It may have confused the Golang syntax with C/C++, where a switch case must end with a \code{break} statement to avoid the fall-through behavior. Golang does not have the fall-through behavior, so the LLM fixed a hallucinated bug that does not exist.

With \ourtool, the number of correct repairs increases to \textbf{276}, which consists of 100 correct repairs for \code{BUG}, 121 for \code{PERFORMANCE}, and 55 for \code{SECURE}. The system now correctly fixes \textbf{69.2\%} of all 399 bugs in the benchmark (was \textbf{40.9\%} with Basic), with a relative improvement of 69.3\%. \figref{fig:eval-rq3} further visualizes the increase in correct repairs by each alert type. We can see that \ourtool increases the number of correct repairs in most cases. In many cases including the \code{BUG.NO.SWITCH.CASE} alert discussed above, the LLM follows the right path to repair and no longer hallucinates when a key example is given. This demonstrates the generalization and practical utility of \ourtool in a real-world scenario.
}

\begin{rqbox}
\textbf{Finding 3:}

\rev{\ourtool generalizes to GoInsight, an imperative static analyzer for Golang, with a 69.3\% increase in the number of correct repairs over the Basic baseline.}
\end{rqbox}

\subsection{Threats to Validity}

\NewDocumentCommand{\rot}{O{20} O{1.5em} m}{\makebox[#2][l]{\rotatebox{#1}{\small{#3}}}}

\begin{table}[tb]
    \centering
    \caption{\rev{Number of correct (successful) repairs against additional 11 recent CVE vulnerabilities.}}
    \begin{tabular}{llllllll}
    \toprule
    Model & \rot{\ourtool} & \rot{Docs} & \rot{Rule Expl} & \rot{Rule Src} & \rot{Similar} & \rot{History} & \rot{Basic} \\
    \midrule
    3.5-Turbo & 5 (6)        & 3 (6) & 3 (4)     & 3 (4)    & 2 (4)   & 3 (3)   & 2 (3) \\
    4o-mini   & 7 (8)        & 5 (6) & 5 (6)     & 6 (7)    & 4 (4)   & 7 (7)   & 5 (6) \\
    \bottomrule
    \end{tabular}
    \label{tab:eval-knowledge-cutoff}
\end{table}

\subsubsection{Internal}
The main internal threat to validity comes from potential data leakage in datasets. To avoid this threat, we drop exact matches of the patched code in our clean code corpus, as described in \secref{sec:impl}.
\rev{
However, there may still be data leakage through LLMs, since some experiment subjects may be used for their training, and we cannot control it. To learn the effect of this threat, we conducted a small additional experiment.

\textbf{Setup:} For the LLM, we used GPT-3.5-Turbo (version 0125, knowledge cut-off in September 2021) and GPT-4o-mini (knowledge cut-off in October 2023). For the benchmark, we used ReposVul~\cite{wang2024reposvul}, which includes C/C++/Java/Python CVE vulnerabilities up to 2023. Following the same process of RQ1-2, we ran CodeQL analysis for recent Java and Python CVE entries between September 2021 and October 2023. As a result, we collected 381 vulnerability-fixing commits, 11 of which fixed a CodeQL alert. We used these 11 cases between the knowledge cut-off of the two LLMs as the benchmark.

\textbf{Result:} The numbers of correct and successful repairs of \ourtool and other baselines are shown in \tabref{tab:eval-knowledge-cutoff}. Although the sample size (11 cases) is insufficient to measure the exact percentage of improvement, we can roughly see that \ourtool increases the number of correct repairs over Basic by at least 2, regardless of whether the knowledge cutoff date is before (i.e., GPT-3.5-Turbo) or after (i.e., GPT-4o-mini) these vulnerabilities. Therefore, this threat should have a limited chance of affecting previous findings.
}

\subsubsection{External}

The main external threat to validity comes from the generalization of our approach: can \ourtool show similar effectiveness in other programming languages, LLMs, experiment subjects, or static analyzers? To address the former two threats, our evaluation has used six LLMs from different vendors with different sizes, and RQ2 also has a breakdown of effectiveness in each programming language. For the latter two threats, RQ3 performs an additional evaluation on GoInsight, an imperative Golang static analyzer, where \ourtool shows a non-trivial effectiveness increase. These experiments reduced the threat of generalization.

\section{Related Work}
\label{sec:related}

\subsection{Program Repair Targeting Static Analysis}

\ourtool shares similar objectives with many works aimed at automatically repairing alerts identified through static analysis alerts.
\citet{liu2021mining} analyzed the distribution of bug-fix pairs for FindBugs violations, and proposed a neural network to automatically identify fix patterns through historical patches.
Tools such as Avatar~\cite{liu2019avatar}, Phoenix~\cite{bavishi2019phoenix}, and Getafix~\cite{bader2019getafix} can cluster and generalize fix patterns for different alert types and use them to generate new patches.
These approaches relied on a large corpus of historical patches for learning fix rules.
\secref{sec:motivation-baseline-rag} and the unsatisfactory effectiveness of the ``History'' baseline in RQ2 have shown that finding only a few high-quality historical patches is already challenging, not to mention building a corpus of historical patches.
In contrast, \ourtool requires only clean code examples and expands the search scope to find high-quality examples with \keypredicates.

Some approaches focus on specific error types, such as memory errors~\cite{gao2015safe,lee2018memfix,hong2020saver,lee2022npex}, API misuse~\cite{cruz-carlon2022patching}, and missing sanitizers or guards~\cite{jain2023staticfixer}.
These approaches are tightly integrated with one or a few analysis rules and are tailored to fix alerts on only these rules.
Approaches such as SpongeBugs~\cite{marcilio2019automatically} and Sorald~\cite{etemadi2023sorald} fix multiple types of alerts by manually specifying a rule for each type.
\ourtool, however, applies to general static analyzers without manual efforts on each alert type. 

Furthermore, some symbolic approaches leverage static analysis information for repair.
Senx~\cite{huang2019using} uses symbolic execution to patch a program to satisfy the user-specified safety properties.
SymlogRepair~\cite{liu2023program} encodes candidate patches as Datalog values and uses constraint solving to efficiently find patches that satisfy Datalog rules.
EffFix~\cite{zhang2023patch} uses incorrectness separation logic to group patches according to their semantic effects and measure how close a patch is to fixing the bug.
These approaches require precisely specifying the repair condition, and face the scalability issue of symbolic methods on a large codebase.
In contrast, \ourtool can be applied to a wider scope of alerts and programs where symbolic approaches are infeasible. Therefore, our approach is complementary to these approaches in handling more alerts and programs.

There is growing interest in deep learning methods for program repair. TFix~\cite{berabi2021tfix} fine-tuned a T5 model to fix ESLint violations. DrRepair~\cite{yasunaga2020graphbased} trained a graph neural network on graphs that connect source codes with diagnostic messages to model the reasoning behind program analysis. Compared with them, \ourtool does not require a training set of historical patches. As discussed above, it is difficult to collect historical patches for many alert types, and thus it is difficult to build a high-quality training set. 

More recently, LLMs have shown great potential in automated program repair~\cite{xia2023automated}.
Approaches such as CORE~\cite{wadhwa2024core} and DeepVulGuard~\cite{steenhoek2024closing} prompt the LLM with information such as the issue description.
The ``Basic'' and ``Docs'' baselines in RQ2 of our evaluation have shown that using only such information in the prompt has limited effectiveness in improving LLMs.

\subsection{Retrieval-Augmented Program Repair}

The performance of large language models can be improved through in-context learning of relevant knowledge~\cite{gao2023what,zhang2024critical}. This motivates some APR approaches using RAG techniques.

FitRepair~\cite{xia2023plastic} exploits the plastic surgery hypothesis by searching for example code snippets in the same project.
Ring~\cite{joshi2023repair} selects examples based on the similarity of the error message.
RAP-Gen~\cite{wang2023rapgen} accounts for both lexical and semantic code matching.
Cedar~\cite{nashid2023retrievalbased} retrieves code based on embedding similarity or frequency analysis.
InferFix~\cite{jin2023inferfix} and VulAdvisor~\cite{zhang2024vuladvisor} retrieve historical patches to enhance the prompt.
VulMaster~\cite{zhou2024out} incorporates code syntax trees and CWE expert knowledge in prompt design and uses the Fusion-in-Decoder technique to deal with the context length limit.
T-RAP~\cite{liu2024trap} uses edit scripts as templates to retrieve similar fixes.
Unlike these RAG approaches, \ourtool considers the static analyzer as a white box and takes advantage of the predicate information in the retrieval process, leading to high-quality \keyexamples and a better effectiveness than the ``Similar'' baseline as shown in RQ2.
\section{Conclusion}
\label{sec:conclusion}

This paper presented \ourtool, a retrieval-augmented generation (RAG) approach that fixes static analysis alerts in program code. The novel insight behind our approach is to identify and utilize \keypredicates from the analysis rule and \keyexamples in a clean code corpus. Based on this insight, we proposed an algorithm to automatically retrieve \keyexamples as the source of demonstration. In this way, we can provide relevant knowledge to the large language model and thus increase its effectiveness, particularly on less common alerts where the model often hallucinates.

We implemented and evaluated \ourtool on multiple static analyzers, programming languages, and LLMs. The evaluation confirmed that \ourtool can increase effectiveness by 27.1\% to 69.3\%, significantly outperforming other baseline RAG and prompt engineering techniques.

\textbf{
The artifacts of this paper are available at FigShare: \href{https://doi.org/10.6084/m9.figshare.26956228}{https://\allowbreak{}doi.org/10.6084/m9.figshare.26956228}.}

\begin{acks}
This work was sponsored by the National Key Research and Development Program of China under Grant No. 2022YFB4501902, the National Natural Science Foundation of China (NSFC) under Grant No. W2411051, and the ZTE Industry-University-Institute Cooperation Funds under Grant No. HC-CN-20210319008.
\end{acks}


\bibliographystyle{ACM-Reference-Format}
\bibliography{bib/merged}

\end{document}